\documentclass[11pt]{article}
\usepackage[utf8]{inputenc}
\usepackage{amssymb,amsmath,amsthm}
\usepackage{graphicx}
\usepackage{cite}
\usepackage{algorithm,algcompatible}

\algnewcommand\algorithmicto{\textbf{to}}
\algnewcommand\RETURN{\State \textbf{return} }

\newtheorem*{remark}{Remark}

\title{WKB Across Caustics: The Screened-WKB Method} \author{Oscar
  P. Bruno\footnote{Computing and Mathematical Sciences, Caltech,
    Pasadena, CA 91125, USA} \and Martin D. Maas$^*$}

\begin{document}
\date{}

\maketitle

\begin{abstract}
  We present a new methodology, based on the WKB approximation and
  Fast Fourier Transforms, for the evaluation of wave propagation
  through inhomogeneous media. This method can accurately resolve
  fields containing caustics, while still enjoying the computational
  advantages of the WKB approximation, namely, the ability to resolve
  arbitrarily high-frequency problems in computing times which are
  orders-of-magnitude shorter than those required by other algorithms
  presently available. For example, the proposed approach can simulate
  with high accuracy (with errors such as e.g. 0.1\%--0.001\%) the
  propagation of 5 cm radar signals across two-dimensional
  configurations resembling atmospheric ducting conditions, spanning
  hundreds of kilometers and millions of wavelengths in electrical
  size, in computing times of a few minutes in an ordinary CPU
  workstation.
\end{abstract}

\section{Introduction}

Computations of high-frequency wave propagation through inhomogeneous media play a pivotal roles in diverse fields such as telecommunications, remote sensing, seismics, quantum mechanics, and optics. A wide range of methodologies have been developed over the last century for the treatment of high-frequency volumetric-propagation problems. Given that direct numerical simulation of the configurations of interest, which comprise thousands to millions of wavelengths in acoustical/electrical size, is unfeasible in 2D and even more in 3D, the proposed approaches usually contain a combination of analytic and numerical approximations. 

The celebrated WKB approximation, also known as the Wentzel-Kramers-Brillouin approximation \cite{bornandwolf, keller1962}, was the first method to obtain accurate solutions to problems involving propagation over large distances, and is based on the introduction of a system of ray-coordinates, over which the amplitude and phase of the solution exhibit slow variations. However, the WKB approximation can break down in certain situations, particularly when the ray mapping becomes singular (i.e. at caustics). Many approaches have been proposed over time to overcome this limitation, most notably the KMAH-index theory, according to which a correction can be incorporated after the caustic, of the form $(-i)^m$, where the constant $m$ depends on the number and type of caustics that the ray has traversed. This formulation still breaks down at caustics, is inaccurate near caustics, and, given its complexity, it is seldom used in practice.

Another notable approach to solve these type of problems is provided by the parabolic approximation introduced in~\cite{leontovich1946}, together with its many subsequent versions and improvements, including the wide-angle approximation~\cite{hardintappert1973}. The method of phase-screens \cite{wu1994}, in turn, which assumes a constant refractivity profile along each vertical volumetric screen, is applicable for certain restricted sets of configurations. While, unlike the classical WKB approach, these methods are valid at and around caustics, their limitations arise from its computational cost. For example, mesh-sizes of the order of $\Delta z \approx \frac{\lambda}{4}$ and $\Delta x \approx 12.5 \lambda$ are reported for propagation distances in the order of a few hundreds of wavelengths (see \cite{jensen2011} and references therein). (Here $z$ and $x$ denote the vertical and range variables, respectively.) The parabolic equation methods are most often based on use of finite-difference approximations, which gives rise to associated dispersion errors, while Fourier expansions in the vertical axis are only applicable in the lowest-order parabolic approximations. The combined effect of large propagation distances, and the presence of dispersion error, and the requirement of fine spatial discretizations, can lead to extremely large computational cost, under reasonable error tolerances, for challenging configurations commonly arising in applications. 

Other notable approaches include the Gaussian beams formulation, with contributions spanning from the 60's, including~\cite{babichbuldyrev1991, vcerveny1982computation,   tanushev2009gaussian,hormander1963} among many others. This formulation is based on an additional approximation to WKB, which seeks to obtain the phase in the form of a quadratic polynomial, whose Hessian matrix is evolved along the ray. This approach eliminates ray-bunching at caustics, and produces fields which remain bounded. However, theoretical convergence as $k \to \infty$ has not been established and is believed to be slow. Moreover, the initial beam representation is a challenging optimization problem, which leads to errors of a few percent even for propagations distances of the order of a small number of wavelengths~\cite{tanushev2009gaussian}.  

An additional approach, known as Dynamic Surface Extension (DSE, see \cite{steinhoff2000new, ruuth2000fixed}), can successfully propagate wavefronts in a Cartesian discretization. However, the amplitude computations present the same limitations as the classical WKB approximation. Finally, the Kinetic Formulation \cite{engquist2003computational} views each ray tracing equation as describing the motion of a "particle” (e.g. photon, phonon). This method presents severe computational difficulties, as the initial conditions and solutions are given in terms of Wigner measures, a $\delta$-function that vanishes for incorrect directions $p$.\looseness = -1  The approach put forth in this paper, on the other hand, is based on the WKB approximation, and overcomes the limitations posed by caustics by resorting to a family of curves (or screens) on which the total field is decomposed in Fourier modes. Each mode is then propagated for large electrical distances (i.e. 20,000$\lambda$ in the example considered in Figure~\ref{multiple-caustics}) which are also short enough that the presence of caustics is avoided for each Fourier mode.  

\section{The Screened-WKB Method}

\begin{figure}
  \begin{center}
    \includegraphics[width=0.45\linewidth]{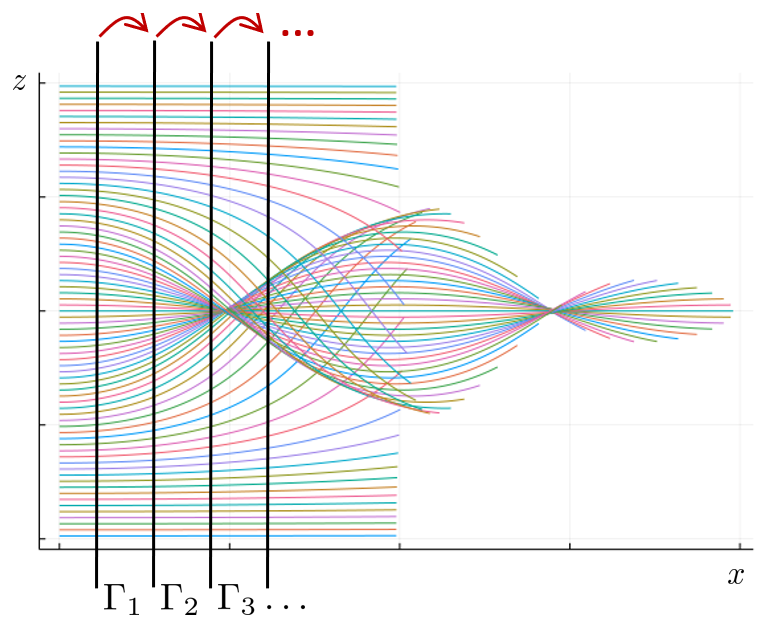}
  \end{center}
  \caption{Schematics underlying the proposed S-WKB method. Here and
    throughout this paper flat screens $\Gamma_q$ are used, but curved
    screens could alternatively be utilized, if
    convenient.\label{S-WKB_screens} }
\end{figure}

We consider, as a model problem, the scalar Helmholtz equation
\begin{align}\label{helmholtz_3D}
 \Delta u(\mathbf{r}) + k^2 \varepsilon(\mathbf{r}) u(\mathbf{r}) = 0 \\
 u = u^s + u^\mathrm{inc} \\
 \lim_{r\to\infty} r\left( \frac{\partial u^s}{\partial r} - iku^s \right) = 0,
\end{align}
whose character at high frequencies presents challenges often found in
diverse fields, such high-frequency electromagnetism, acoustics,
seismics and quantum mechanics.  

The proposed screened-WKB approach first introduces a family of curves (or screens) $\Gamma_q$, for $q=0,1,\dots,N_s$, as depicted in Figure~\ref{S-WKB_screens}. The method proceeds by propagating the solution from one screen to the next on the basis of Fourier expansions on the screens $\Gamma_q$ and applications of the classical WKB approach for each separate Fourier mode.

For conciseness, we consider planar screens of the form
$\Gamma_q = \lbrace (x_q,z) : z\in(z_a, z_b) \rbrace$. The initial
conditions on $\Gamma_0$ are user-prescribed, and given by:
\begin{equation}\label{intial_cond}
    u |_{\Gamma_0} = u^\mathrm{inc} |_{\Gamma_0}
\end{equation}

As we seek to employ the classical WKB approximation between $\Gamma_q$ and $\Gamma_{q+1}$ for each $q$, we seek to represent the incident field on each screen $\Gamma_q$, arising from propagation of the field from $\Gamma_{q-1}$ (or given by~\eqref{intial_cond} for $q=0$) by expressions of the form
\begin{equation}\label{ray_bundle}
 u (x,z) \approx \sum_{n=-N/2}^{N/2-1} A^q_n(x,z) \exp(ik \psi^q_n(x,z)).
\end{equation}
In other words, this expression corresponds with sums of WKB-expansions.

\subsection{The $z$-periodic case}

We will first introduce the proposed approach under the assumption that the incident field and the solution are $z$-periodic functions, which will enables the construction of a simple algorithm and provide some numerical results. Under such hypothesis, the method can begin with a ``vertical'' DFT, and provide a Fourier representation that is exploited to . In subsequent sections, a generalization based around the Fourier Continuation strategy will be presented, that is noy only more general but also computationally more efficient, along with specific methods to deal with homogeneous Dirichlet and Neumann boundary conditions.

\subsubsection{From FFTs to WKB expansions}

Assuming the incident field and the solution are periodic in $z$ (In particular, this also includes cases wherein the solution decays rapidly outside a bounded interval in the $z$ variable.), for a given screen $\Gamma_q$, a ``vertical'' DFT
can be used by introducing an equi-spaced grid
\begin{equation}
  \lbrace z_m\,:\,  m=-N/2,\dots,N/2-1 \rbrace
\end{equation}
in the interval $[z_a,z_b]$. Performing an FFT yields
\begin{equation}\label{fft_eq}
 w^q_j = \sum_{m=-N/2}^{N/2-1} u(x_q,z_m) e^{-i j z_m}.
\end{equation}
Then, we can re-express the field $u(x_q,z)$ in terms of an inverse DFT:
\begin{equation}\label{ifft_eq}
u |_{\Gamma_q} \approx  \frac{1}{N} \sum_{m=-N/2}^{N/2-1} w^q_j e^{i j z_m},
\end{equation}
Now, by introducing the functions $\psi^q_n(x_q,z)$ defined by 
\begin{equation}\label{fourier_psi}
 \psi^q_n(x_q,z) = \left( z + \frac{z_b-z_a}{2} \right)\frac{2 n\pi}{ k(z_b-z_a) }.
\end{equation}
we obtain an expression of the form
\begin{equation}\label{ifft_eq_phi}
  u |_{\Gamma_q} \approx  \frac{1}{N} \sum_{m=-N/2}^{N/2-1} w^q_j e^{i j \psi^q_n(x_q,z_m)},
  \end{equation}
which in turn, corresponds to an expression of the form \eqref{ray_bundle} evaluated in a discretization of the points of the screen $\Gamma_q$.

So far the proposed method is physically-exact. Under the assumptions that the incident field propagates mostly in the $x$ direction, the resulting fields should be slowly-varying in the $z$ direction, which makes the FFT approach viable. However, the problem of propagating each term in the expansion \eqref{ifft_eq_phi} up to the next screen $\Gamma_{q+1}$, involves highly oscillatory functions. In order to do this, we seek asymptotic solutions to equation \eqref{helmholtz_3D}. In a procedure which is equivalent to the WKB expansion (see e.g. \cite[chapter 3]{jensen2011}), we consider trial solutions of the form
\begin{equation}\label{trial_sol}
u(r) = A(r) \mathrm{exp}\left[ik \psi(r) \right]
\end{equation}
and expand the amplitude in inverse powers of the wavenumber, 
\begin{equation}
 A(r) = A_1(r) + \frac{A_2}{ik} + \frac{A_3}{(ik)^2} + \dots
\end{equation}

Replacing the trial solution \eqref{trial_sol} in the Helmholtz \eqref{helmholtz_3D} results in 
\begin{align}
 \Delta u(r) + k^2\varepsilon(r) u(r) 
 & = k^2 \left[ \varepsilon(r) - \psi(r) \right] \\
 & + ik \left[ 2 \nabla \psi \cdot \nabla A_1 + A_1 \Delta \psi \right] \\
 & + \Delta A
\end{align}

Up to degree 2 in $k$, we obtain the Eikonal equation and the transport equations:
\begin{align}
  (\nabla \psi)^2 = \varepsilon(r) \label{first_order_system_1} \\
  2\nabla \psi \cdot \nabla A + A \Delta \psi = 0 \label{first_order_system_2}
\end{align}

\subsubsection{From the $\Gamma_q$ screen to $\Gamma_{q+1}$ via WKB \label{wkb_propagation}}

The algorithm proceeds by propagating each mode in the expansion \eqref{ifft_eq_phi} by means of WKB expansions. In other words, in order to perform this asymptotic procedure up to order $O(\frac{1}{k^2})$, we must solve the eikonal and transport equations \eqref{first_order_system_1} and \eqref{first_order_system_2} for each mode. Interestingly, the proposed approach is compatible with any solution strategy for this problem. 

For completeness, in the present section we discuss a high-order solver based on characteristics tracing with ODE solvers and piecewise interpolation.

As a first step, the system of characteristics equations for the system \eqref{first_order_system_1} and \eqref{first_order_system_2} (obtained in Appendix \ref{wkb_characteristics}) is solved by employing numerical ODE solvers.

In order to obtain initial conditions for the eikonal equation \eqref{first_order_system_1} for each mode of \eqref{ifft_eq_phi} on
$\Gamma_q$, we resort to \eqref{fourier_psi}, while the initial conditions for the transport equation \eqref{first_order_system_2} are given by the Fourier amplitudes $w^q$. We then have the following set of initial conditions
\begin{eqnarray}
 \partial_z \psi_n(x_q,z) = \frac{2 n\pi}{ k(z_b-z_a) } \\
 \partial_x \psi_n(x_q,z) = \sqrt{ \varepsilon(x_q,z) - \left(\frac{2 n\pi}{k(z_b-z_a)}\right)^2 } \\
 A(x_q,z) = w^q
\end{eqnarray}

Applying an ODE solver to the system of characteristics (equations \eqref{charsys} and \eqref{charsys_s} in the Appendix \ref{wkb_characteristics}), yields a finite number of adequately spaced geometrical-optics rays, and corresponding values of $\psi_n$ and $A_n$ along the rays for the $n$-th mode. 

Standard interpolation procedures can then be used on $\Gamma_{q+1}$ to obtain approximate values of $u$ on the 1D Cartesian grid $(x_{q+1},z_m)$ ($-\frac N2\leq n\leq \frac N2 -1$) on $\Gamma_{q+1}$:
\begin{equation}\label{ray_bundle_q_plus_1}
  u(x_{q+1},z_m) \approx \sum_{n=-N/2}^{N/2-1} A^q_n(x_{q+1},z_m) \exp(ik \psi^{q}_n(z_m)).
\end{equation}

The next iteration of the algorithm can then be initiated by expanding $u(x_{q+1},z)$ in a Fourier series along $\Gamma_{q+1}$. Repeating this procedure for all screens, the field $u$ over the domain of interest can be obtained. When desired, additional interpolation points can be added for visualization purposes.

\begin{remark}
  Importantly, by adequately selecting the spacing of the screens $\Gamma_q$, it can ensured that all the modes $-\frac N2\leq n\leq \frac N2 -1$ propagate to the next screen $\Gamma_{q+1}$ without incurring caustics. 
\end{remark}

\subsubsection{Summary of the algorithm for the $z$-periodic case}

The overall algorithm can thus be summarized as follows:
\begin{algorithm}
  \caption{Screened WKB}
  \label{alg_init}
\begin{algorithmic}
  \STATE Expand $u_{inc}$ with a vertical FFT \eqref{fft_eq}, to obtain \eqref{ray_bundle}.
  \STATE Solve Eikonal and transport equations for each $\psi_n, A_n$.
  \STATE Interpolate $\psi_n, A_n$ to a Cartesian grid.
  \STATE Perform sum over $n$ to obtain new incident field on next screen.
\end{algorithmic}
\end{algorithm}

\begin{remark}
  The method described for the periodic case, is not only limited to such configurations, but it also has a computational cost of $O(n^2)$, where $n$ is the number of points required for the discretization of the $z$ variable. This factor can be the source of an important limitation for problems that are large enough in the vertical dimension.
  \end{remark}

\subsection{Overlapping Fourier-Continuation SWKB \label{overlapping_grid}}

We now present a generalized method, based on an overlapping Fourier
Continuation decomposition, that produces representations of similar
form to~\eqref{ray_bundle}, which is applicable to periodic and
non-periodic functions.  As a first step, in this section we consider
the case in which the incident field $u_\mathrm{inc}$ and resulting
solution between two consecutive screens vanishes in a neighborhood of
the domain boundary; the subsequent treatment of boundary conditions
is discussed in Section~\label{bc}.

We thus consider an overlapping-grid decomposition of the interval
$[z_a,z_b]$, that is, a collection of ``overlapping patches'' of the
form $[a_p,b_p] \subset [z_a,z_b]$ with $a_{p+1}< b_{p}]$, together
with a uniform grid $\lbrace z_m^p : 0 \le m \le M \rbrace$ defined on
each of those patches. The method proceeds by obtaining a Fourier
decomposition on each patch by means of the Fourier Continuation
method~\cite{amlani2016fc,bruno2010high}, which produces accurate
Fourier expansions of non-periodic smooth functions---which play a
similar role to that of \eqref{ifft_eq} in the periodic context. As the
FC method is applied in each patch, we obtain, in the overlapping
regions, two simultaneous representations of the same function. In
order to obtain a single representation which remains smooth throughout
$[z_a,z_b]$, we employ a smooth partition of unity which vanishes at
the endpoints of each patch. This process results in a representation
that closely resembles \eqref{ifft_eq_phi}:
\begin{equation}\label{ifft_eq_phi_patches}
  u |_{\Gamma_q} \approx   \sum_{p=0}^P  S^p(z) \sum_{m=-N/2}^{N/2-1}\frac{w^{q,p}_j}{N} e^{i j \psi^{q,p}_n(x_q,z_m)},
\end{equation}
In particular, we can now apply the classical WKB expansion to each
Fourier mode at each patch, as previously done in the periodic case,
by considering initial conditions $\frac{S^p(z)}{N} w^{q,p}_j$ for the
the amplitud terms in the transport equation $A$.

\subsection{Treatment of non-periodic boundary conditions\label{bc}}

The continuation methodology put forth in previous sections enables
the treatment of non-periodic boundary conditions. In this section, we
present methods to deal with either homogeneous Dirichlet or Neumann
conditions.  The proposed approach is based on constructing a solution
in an extended domain and later reflecting it in such a way that the
boundary condition is satisfied. As a first step, we consider a
``reflected'' permittivity $\tilde{\epsilon}(r)$ along the boundary,
given by
\begin{equation}
  \tilde{\epsilon}(x,z) = 
  \left\lbrace
  \begin{array}{cl}
    \epsilon(x,z) & z \in [z_a,z_b] \\
    \epsilon(x,2 z_b-z) & z > z_b \\
    \epsilon(x,2 z_a-z) & z < z_a
  \end{array}
  \right .
\end{equation}
After a solution $u_{tot}$ is produced in the extended domain, we can obtain a solution for the Dirichlet (resp. Neumann) problem by defining
\begin{equation}\label{u_refl}
  u_\mathrm{refl}(x,z) = u_\mathrm{tot}(x,z) \pm u_\mathrm{tot}(\tilde{x},z)
\end{equation}
where $\tilde{x}=2z_a-z$ in a neighborhood of the endpoint $z_a$, or
$\tilde{x}=2z_b-z$ in a neighborhood of $z_b$.  It is easily verified
that $u_\mathrm{refl}$ satisfies Neumann (resp. Dirichlet) boundary
conditions. However, this reflection procedure also introduces a
singular derivative at each of the endpoints, which, in view of the
Gibbs phenomenon, could lead to large errors in our Fourier
representations. This means that this reflection procedure cannot be
employed in patches that contain the boundaries. This is not a
limitation, as the reflection can be applied to all other patches,
leaving the boundary patches unmodified until all iterations are
completed, to produce the final reflection at the end of the
algorithm.

\section{Numerical results}

In order to evaluate the accuracy of the proposed S-WKB method, we
compare with solutions obtainable by means of separation of variables
(see Section \ref{sl-solver}) for $x$-invariant
permittivity. Importantly, the method described in Section
\ref{sl-solver} is physically-exact---i.e., which contain no
approximations to \eqref{helmholtz_3D} other than those inherent in
the well established numerical solver utilized. For a number of test
cases we use the exponential permittivity model
\begin{equation}\label{gaussian_permittivity}
  \varepsilon(z) = 1 + a e^{-b z^2}
\end{equation}
Together with an incident field given by a Gaussian beam
\begin{equation}\label{inc_field}
u_\mathrm{inc}(x,z) = \int_{-\infty}^\infty e^{i \sqrt{k^2 - \beta^2} x + i \beta z} e^{-\frac{\beta^2}{\sigma^2}} d\beta 
\end{equation}
wherein the integral in the variable $\beta$ is evaluated via standard numerical integration techniques.

\subsection{High-order reference solutions for $x$-invariant permittivity \label{sl-solver}}

In order to assess the accuracy of the proposed approach, reference
solutions obtained for $x$-invariant permittivity
(i.e. permittivity of the form $\varepsilon(x,z) = \varepsilon(z)$)
are used, for which exact solutions to \eqref{helmholtz_3D} may be
obtained via separation of variables:
\begin{equation}\label{series_sol}
u(x,z) = \sum_{i=0}^\infty a_i e^{i \alpha_i x} \phi_i(z).
\end{equation}
Substituting \eqref{series_sol} in \eqref{helmholtz_3D} leads to
\begin{equation}
\sum_{i=0}^\infty (-\alpha^2 + \phi_i''(z) + k^2\varepsilon(z)) a_i e^{i \alpha_i x} = 0.
\end{equation}
Using the orthogonality of the complex exponential functions we then obtain
\begin{equation}\label{alpha_terms}
(-\alpha_i^2 + \phi_i''(z) + k^2\varepsilon(z)) a_i = 0.
\end{equation}
It follows that the non-zero coefficients $\alpha_i$ in
\eqref{series_sol} satisfy the Sturm-Liouville problem:
\begin{equation}\label{phi_equation}
\phi_i''(z) + k^2 \varepsilon(z) \phi_i(z) = \alpha_i^2 \phi_i(z)
\end{equation}
for given boundary conditions on $z$, which depend on the
configuration being considered. The resulting Sturm-Liouville problem
can be discretized with high-order spectral methods. For illustration
purposes, in the present paper we employed the spectral eigensolver
\cite{olver2013fast}, which is available in the ApproxFun.jl Julia
package. In the context of the examples considered in the present
paper either periodic or homogeneous Dirichlet and Neumann boundary
conditions need to be enforced.

A separation-of-variables solution method may be obtained that matches
the series expansion \eqref{series_sol} to given initial conditions at
$x=0$:
\begin{equation}\label{series_init}
  u_\mathrm{inc}(z) = \sum_{i=0}^\infty a_i \phi_i(z).
\end{equation}
Given that the eigenfunctions $\lbrace \phi_i \rbrace$ are orthogonal in $[z_a,a_b]$ we  obtain
\begin{equation}\label{sl_expansion}
  a_i = \frac{\int_{z_a}^{z_b} u_\mathrm{inc}(z) \phi_i(z) dz}{\int_{z_a}^{z_b} \phi_i(z)^2 dz}.
\end{equation}
These integrals can be calculated numerically using suitable
quadrature rules, which in particular are handled adaptively by the
ApproxFun.jl package. Given the coefficients $a_i$, the solution in
the entire domain can be evaluated using~\eqref{series_sol}.

\subsection{Results for the $z$-periodic case}

In our first example we consider the permittivity
model~\eqref{gaussian_permittivity} with parameters $a=10^{-4}$ and $b=10^{-4}$---which, at C-band, results in a configuration that gives rise to a single caustic of cusp type for the first $40$km ($800,000$ wavelengths) in horizontal propagation range. 
\begin{figure}
  \centering
  \includegraphics[width=0.3\linewidth]{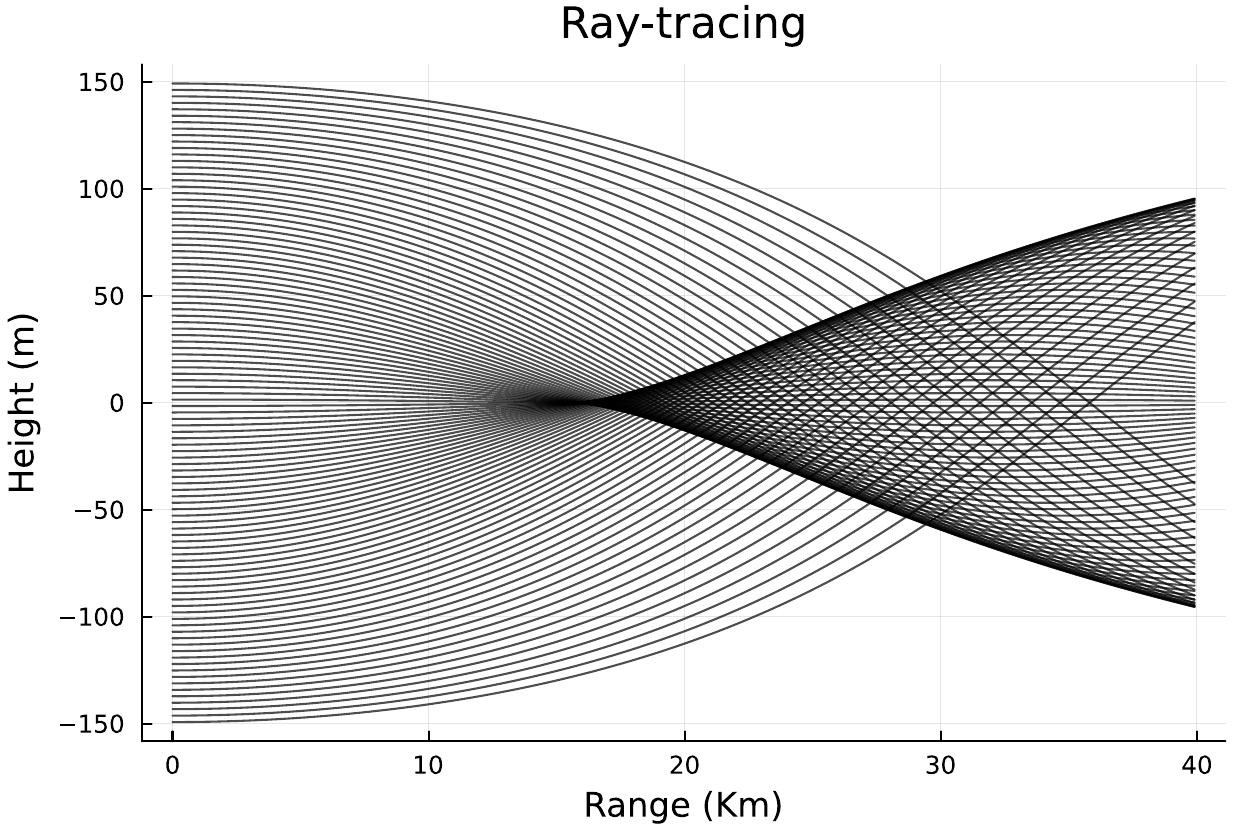}
  \caption{Example 1: Ray-tracing leading to a single cusp
    caustic. \label{single-caustic} }
\end{figure}
The geometrical optics rays for this configuration are displayed in
Fig. \ref{single-caustic}. It can be seen that the configuration
presents a number of caustics: a single point of focusing, together
with an envelope of rays.
\begin{figure}
   \includegraphics[width=0.32\linewidth]{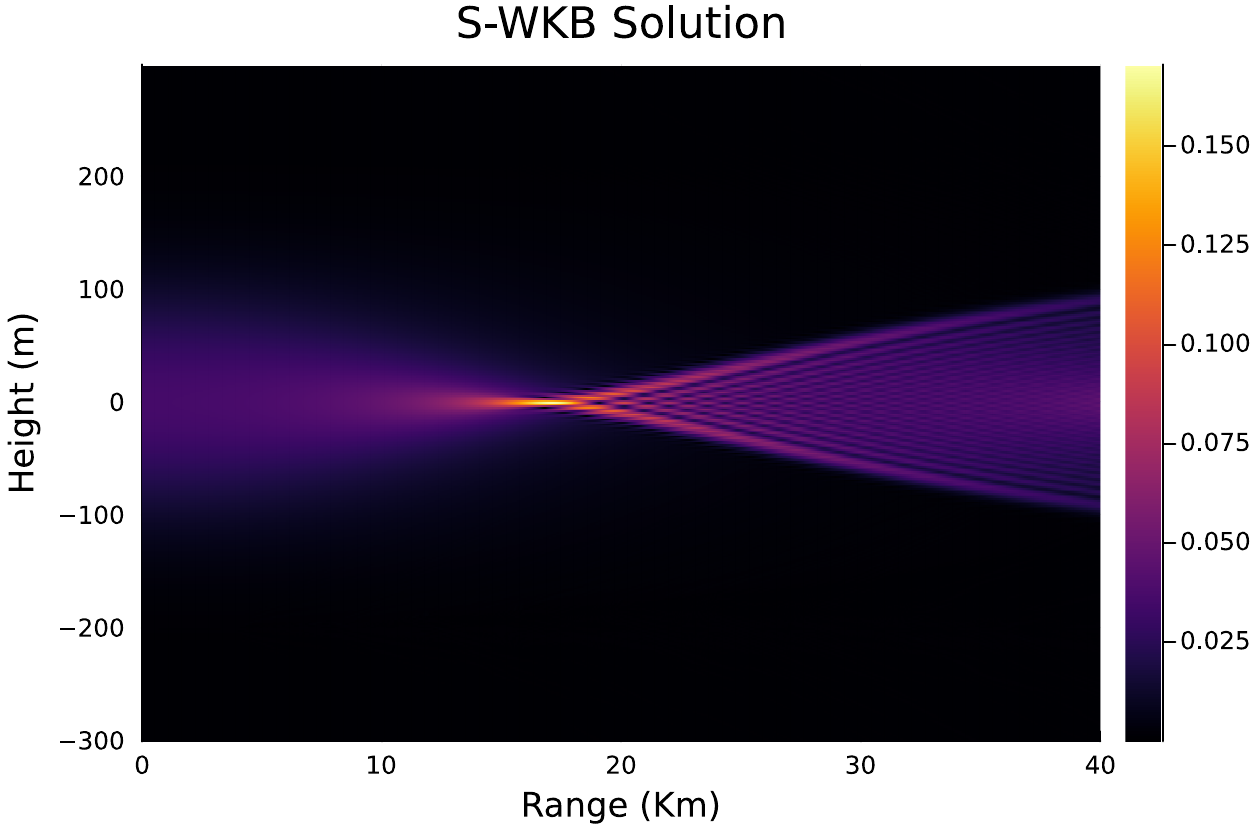}
   \includegraphics[width=0.32\linewidth]{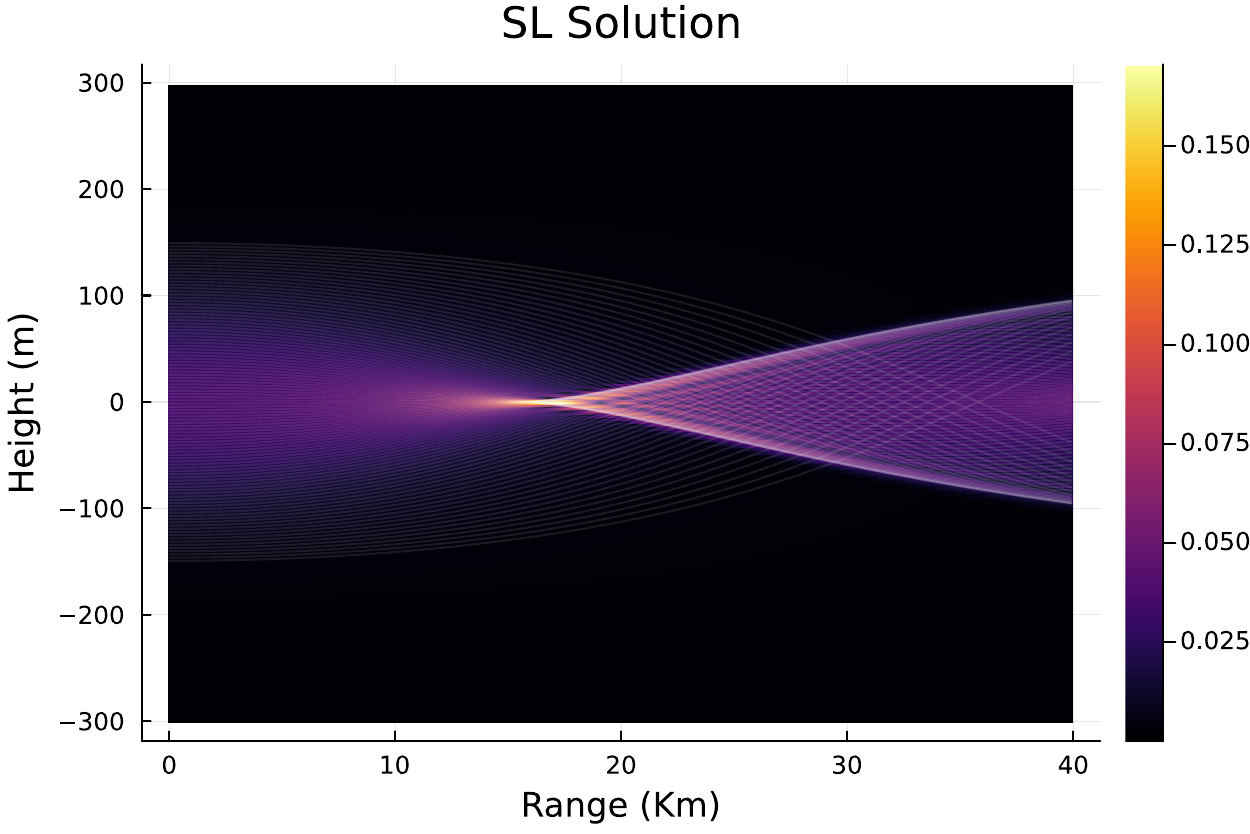}
   \includegraphics[width=0.32\linewidth]{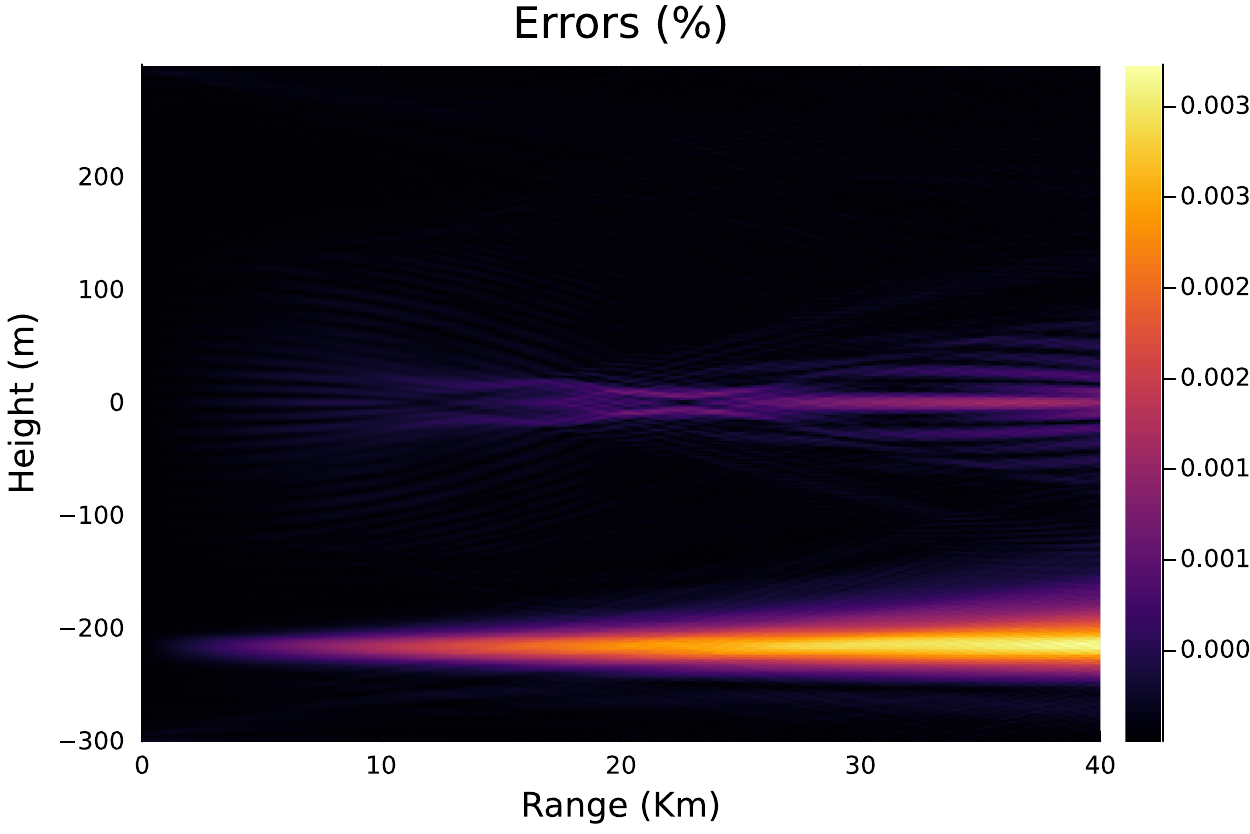}
   \caption{S-WKB solution (left), and physically-exact
     separation-of-variables solution with super-imposed
     geometrical-optics rays (center), with $k=125$, along a propagation
     domain $40$ km ($800,000$ wavelengths) in
     range, and errors (right) for the ``single-caustic'' solution a relative error of the order of $10^{-5}$ was obtained throughout the propagation domain \label{single-caustic-wkb}}
\end{figure}
The S-WKB solution, alongside the Sturm-Liouville solution with
superimposed ray-tracing, and the corresponding relative errors, are
depicted in Fig. As shown, the relative errors for this configuration
are of the order of $10^{-5}$. Employing $400$ Fourier modes and a
total of $40$ screens, the S-WKB solution in this case was obtained in
a computing time of 2 minutes in a single-core.

\begin{figure}
  \centering
  \includegraphics[width=0.5\linewidth]{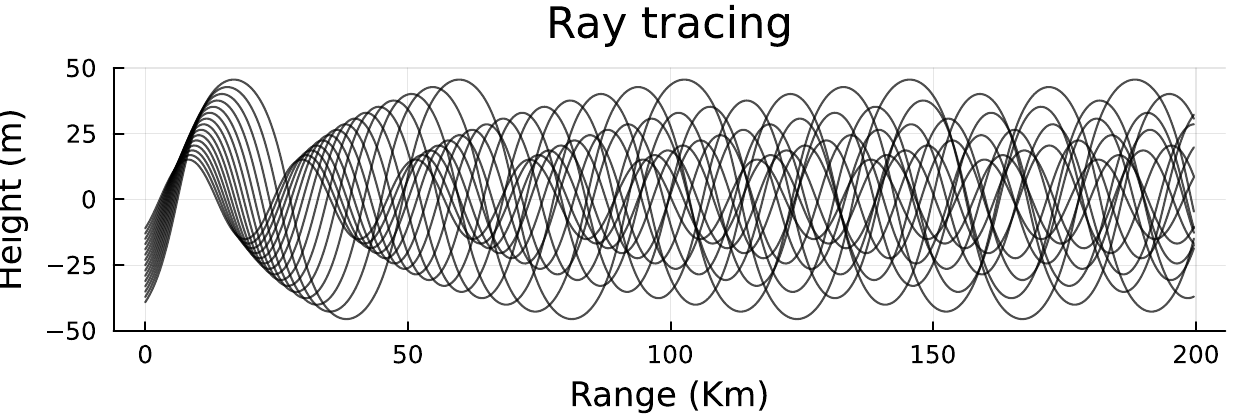}
  \caption{Geometrical optics rays for a ``ducting''
    configuration. \label{ducting-raytracing}}
 \end{figure}

\begin{figure}
  \centering
  \includegraphics[width=0.45\linewidth]{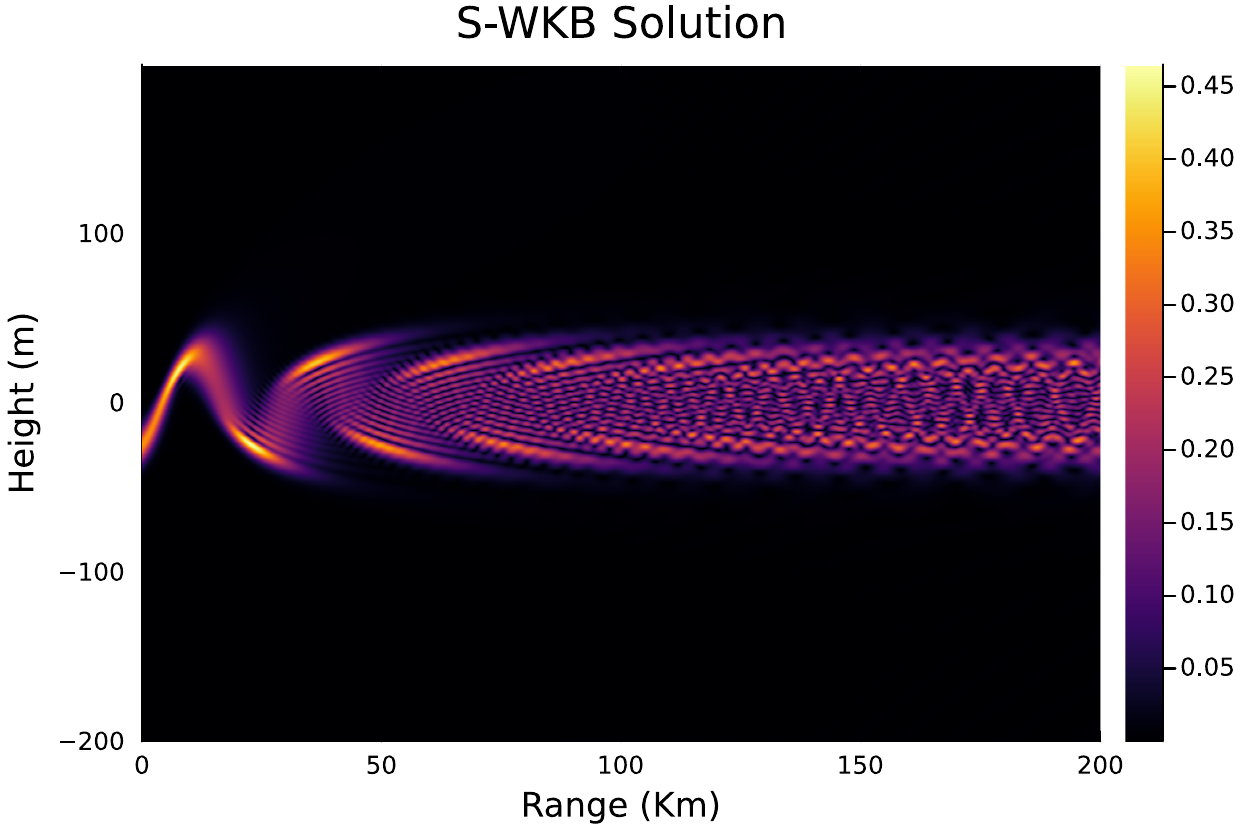}
  \includegraphics[width=0.45\linewidth]{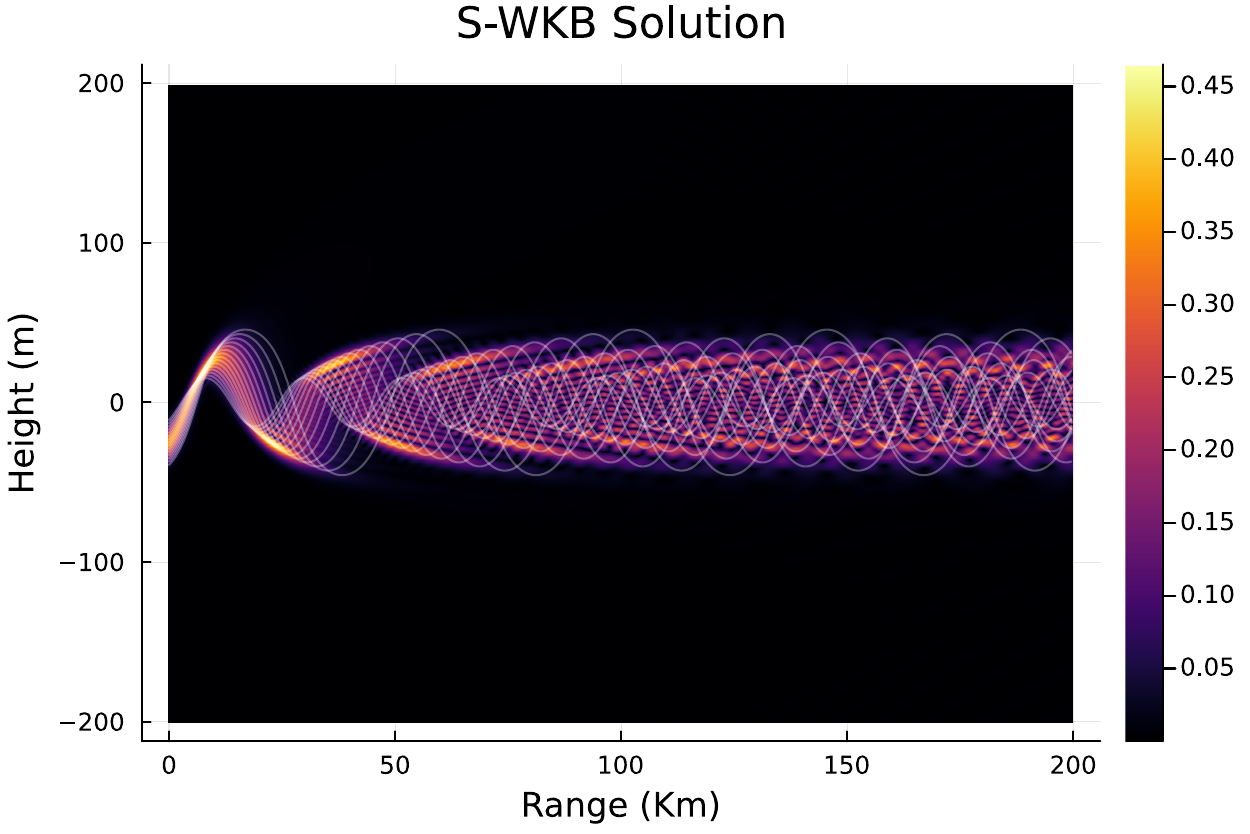}
  \caption{``Multiple caustics'' test case depicting an idealized
    ``ducting'' configuration. S-WKB field values (top), and field
    values with super-imposed geometrical-optics rays (bottom). The
    geometrical optics rays are depicted in
    Figure~\ref{ducting-raytracing}.\label{multiple-caustics}}
 \end{figure}

For our next example we consider a ``ducting'' configuration, in which
the incident Gaussian beam is tilted by an angle of $0.2^\circ$, and
where the Gaussian permittivity~\eqref{gaussian_permittivity} was used
with parameters $a=10^{-4}$ and $b=10^{-3}$---in such a way that the
energy is contained within a bounded interval along the $z$ axis. The
geometrical-optics rays form a complex system with multiple caustics,
as depicted in Fig. \ref{ducting-raytracing}. We consider the
propagation of this signal over a range of $200$km (4 million
wavelengths) in range, which was obtained with an error of $0.1\%$ in a 4 minute single-core computation. 

\subsection{Results for non-periodic cases}

\begin{figure}
  \centering
  \includegraphics[width=0.45\linewidth]{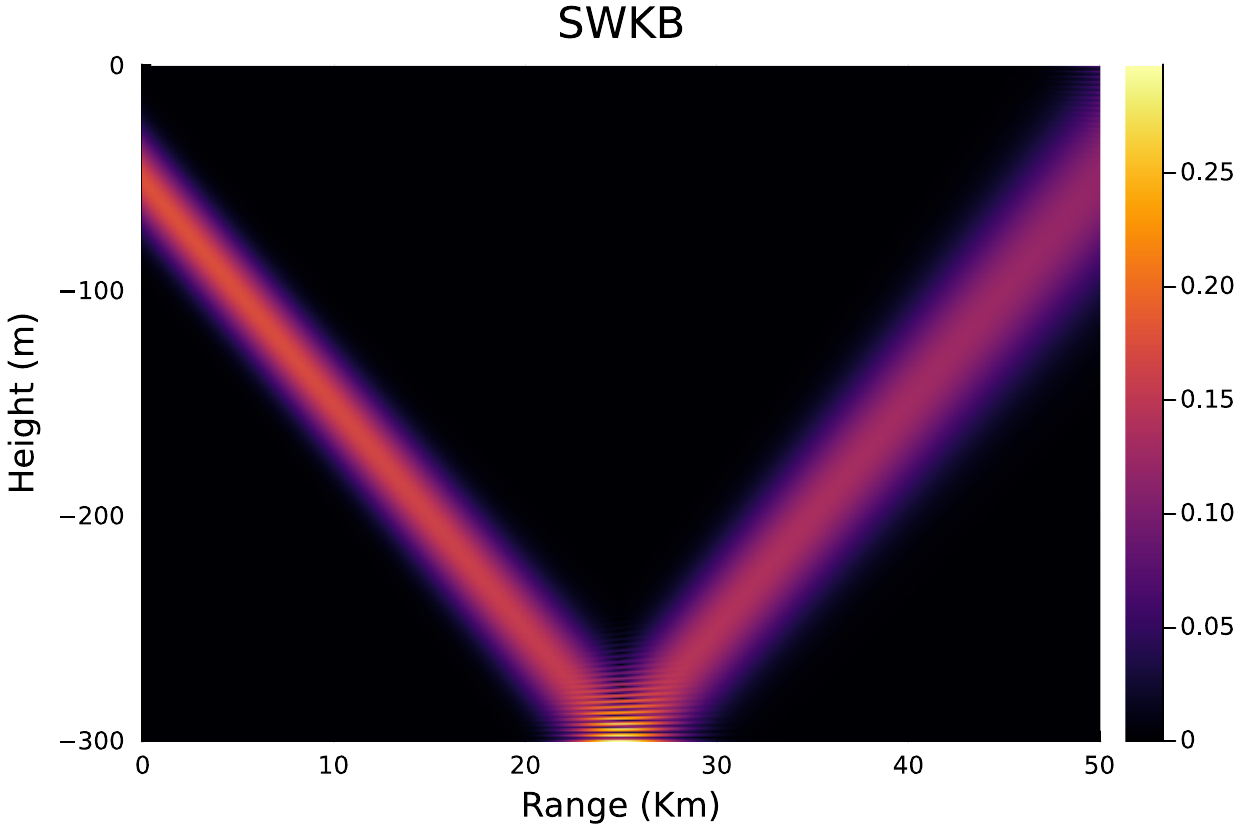}
  \includegraphics[width=0.45\linewidth]{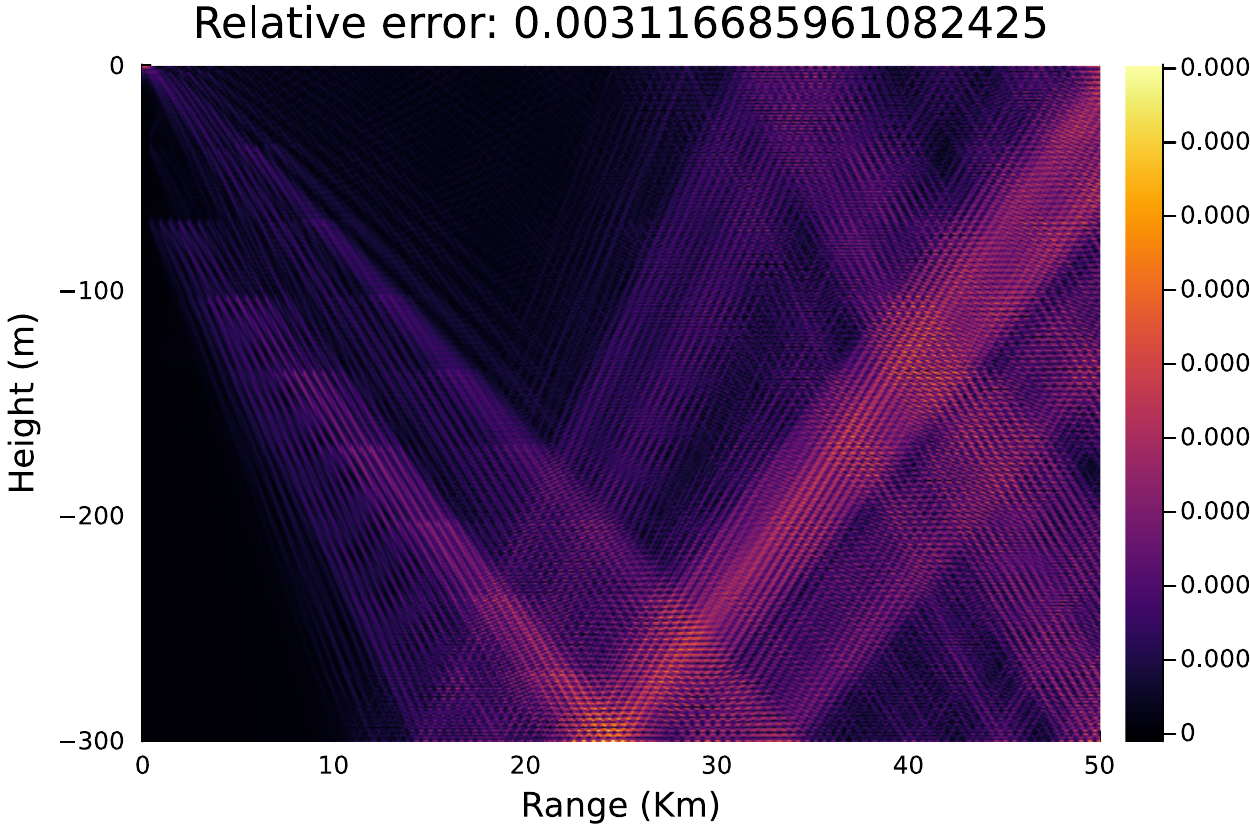}
  \includegraphics[width=0.45\linewidth]{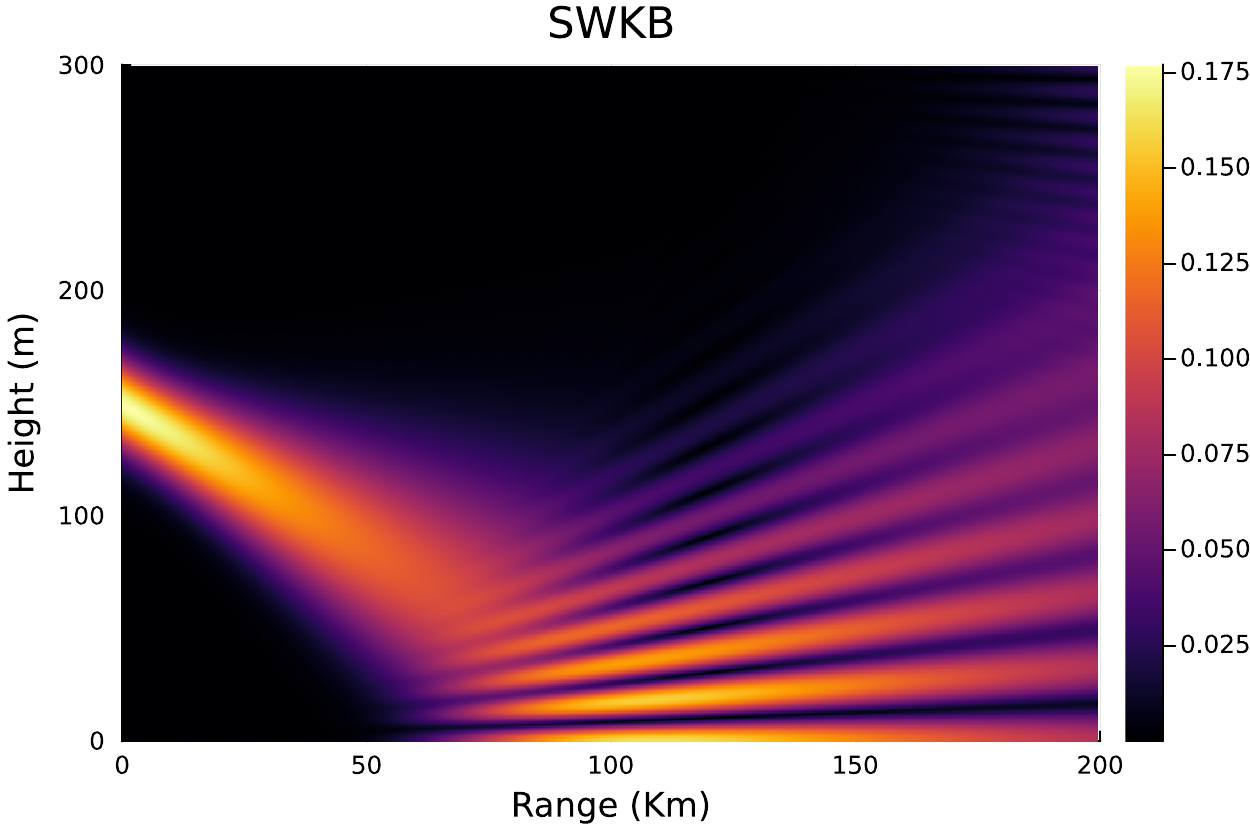}
  \includegraphics[width=0.45\linewidth]{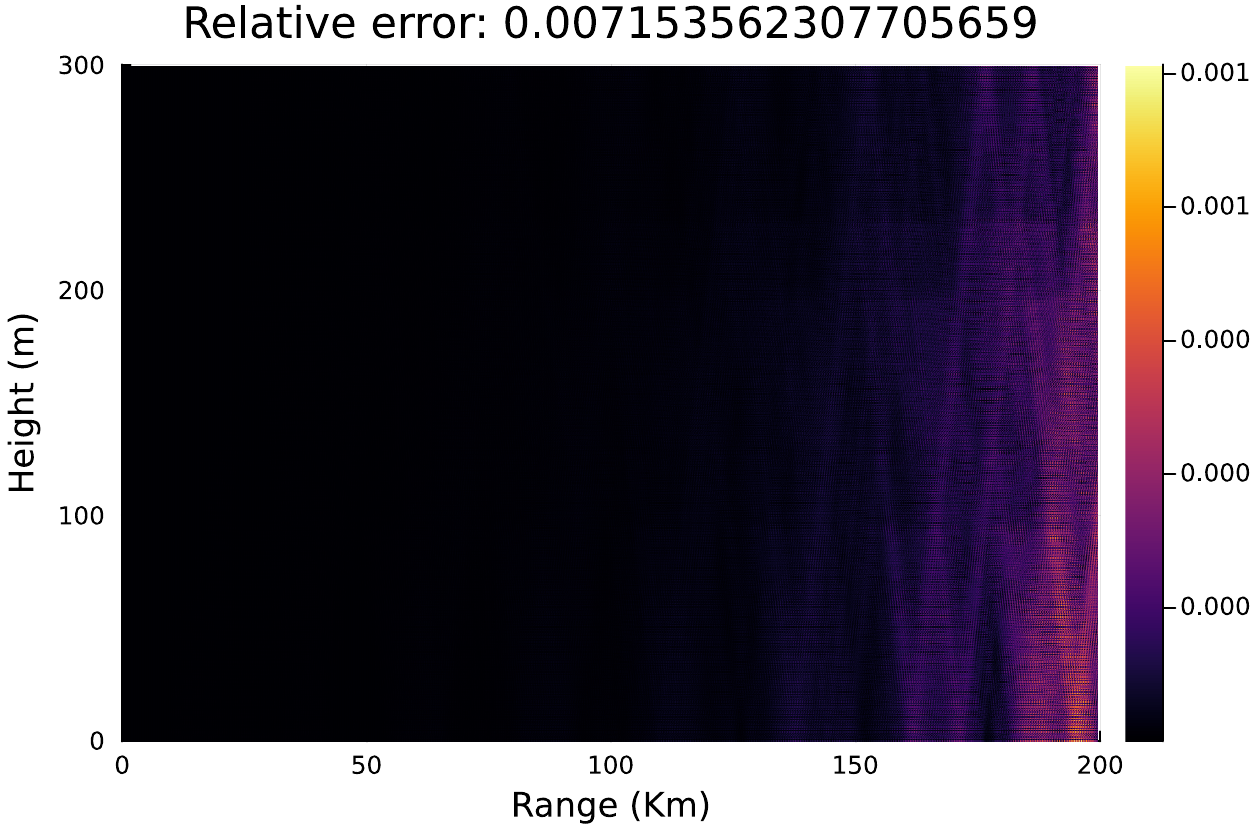}
  \includegraphics[width=0.45\linewidth]{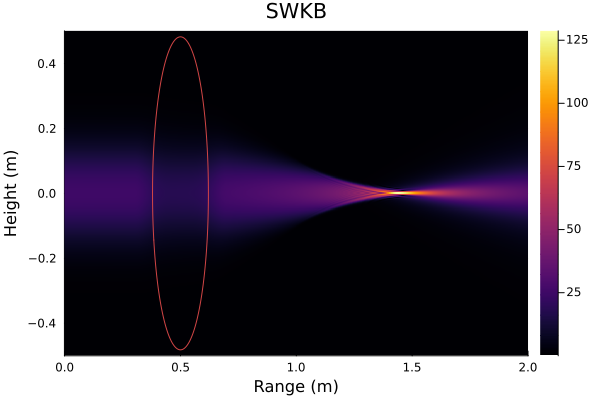}
  \caption{ Beam reflection for a high incidence angle, beam splitting for near-grazing incidence over Neumann boundary conditions, and beam focusing from a permittivity ``lens''. \label{beam-splitting}}
\end{figure}

\begin{figure}
  \centering
  \includegraphics[width=0.45\linewidth]{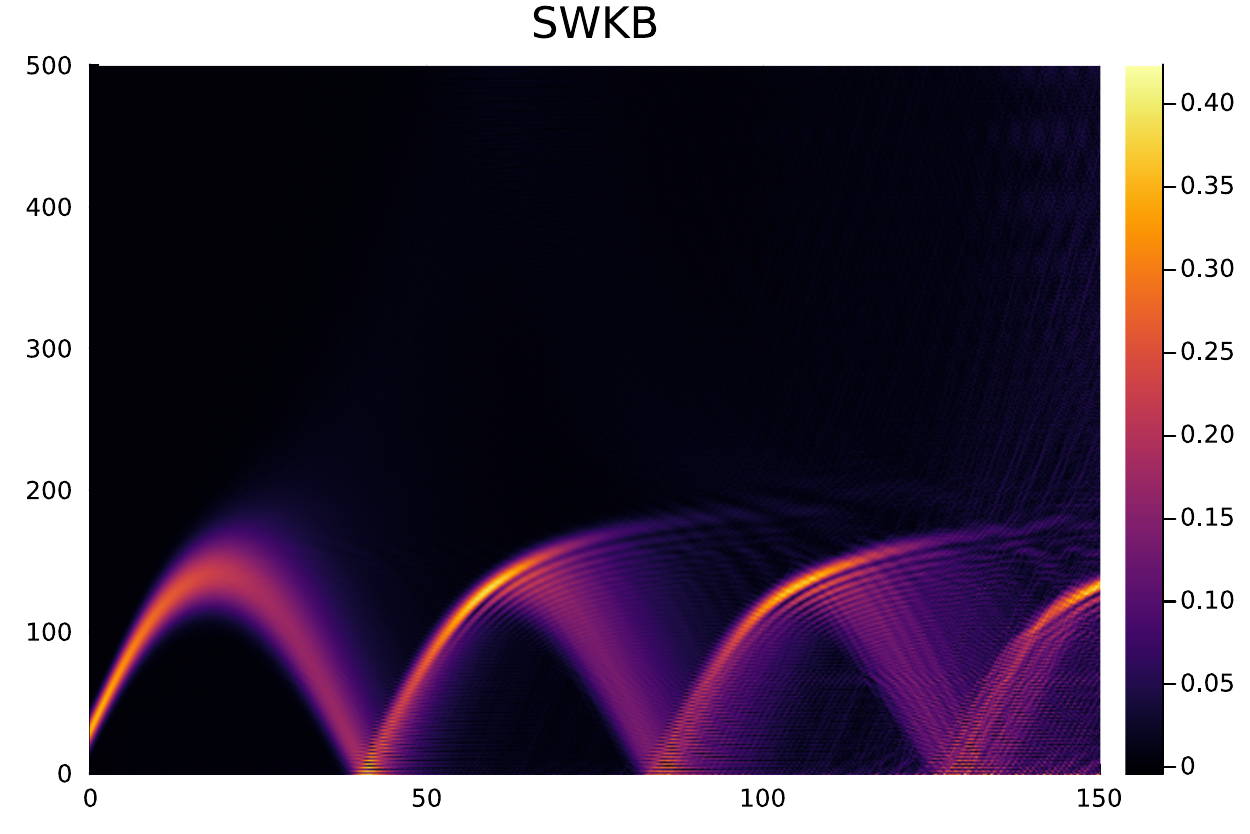}
  \caption{Simplified super-refractive atmospheric conditions \label{super-refractive}}
\end{figure}

The numerical results based on FC representations are similar in
nature to that of the periodic case, with two notable exceptions:
non-periodic boundary conditions may be considered, including boundary
reflections, on one hand; and problems with substantially larger
heights can be tackled, in view of the reduced computational
complexity of the method. The corresponding numerical results are
displayed in Figures~\ref{beam-splitting} and~\ref{super-refractive}

In Fig.~\ref{super-refractive} we consider a simplified version of a super-refractive atmosphere with  permittivity given by
\begin{equation}\label{atmospheric_permittivity}
  \varepsilon(z) = 1.0004 - a e^{-b z^2}
\end{equation}
with $a=1.5 \times 10^{-4}$, $b=1.5 \times 10^{-4}$.

\appendix
\section{The system of characteristics for the eikonal and transport equations \label{wkb_characteristics}}

Given that \eqref{first_order_system_1}-\eqref{first_order_system_2} is a system of first-order equations, it can be solved with the method of characteristics. Following [Garabedian], after introducing the ray coordinates $x(s,t),z(s,t)$, the following system of ordinary differential equations is obtained
\begin{equation}\label{eikonal}
\psi_x^2 + \psi_z^2 - \varepsilon(x,z) = 0 
\end{equation}
results in
\begin{equation}\label{charsys}
    \left\lbrace
    \begin{array}{rl}
        \partial_t x = & 2\psi_x \\ 
        \partial_t z = & 2\psi_z \\
        \partial_t \psi = & 2\varepsilon \\
        \partial_t \psi_x = & \varepsilon_x \\
        \partial_t \psi_z = & \varepsilon_z
    \end{array}      
    \right .  
\end{equation}
where all quantities are evaluated along the characteristics.

The transport equations, upon replacement of \eqref{charsys} in \eqref{first_order_system_2}, yields the additional term
\begin{equation}
    (x_t,z_t) \cdot \nabla A = - \Delta \Psi A
\end{equation}
where the left-hand side can be also expressed as a derivative along the rays
\begin{equation}\label{transport_xt}
    A_t = -\Delta \Psi A
\end{equation}

\subsubsection*{Laplacian on the ray coordinates}
In order to solve the first transport equation, we need to obtain an expression for the Laplacian
$$ \Delta \Psi(s,t) = \psi_{xx}(x(s,t),z(s,t)) + \psi_{zz}(x(s,t),z(s,t)) $$

In the existing literature, this second derivative has been traditionally computed by means of finite differences, which leads to lower orders of accuracy. We introduce an alternative method, based on the system of characteristics \eqref{charsys}. 

We proceed by taking derivatives with respect to $s$ in \eqref{charsys}. Assuming smoothness, we can the interchange the differentiation order to obtain
\begin{equation}\label{charsys_s}
    \left\lbrace
    \begin{array}{rl}
        \partial_t (X_s) = & 2P_s \\ 
        \partial_t (Z_s) = & 2Q_s \\
        \partial_t (\Psi_s) = & 2 \varepsilon_{x}X_s + 2 \varepsilon_{z}Z_s \\
        \partial_t (P_s) = & \varepsilon_{xx}X_s + \varepsilon_{xz}Z_s \\
        \partial_t (Q_s) = & \varepsilon_{zx}X_s + \varepsilon_{zz}Z_s
    \end{array}      
    \right .  
\end{equation}
This ODE system can be integrated if corresponding initial conditions for $(X_s,Z_s,\Psi_s,P_s,Q_s)$ are provided. 

To obtain an expression for $\Delta \Psi$, we employ the relations
\begin{equation}
  \left\lbrace
  \begin{array}{rl}
   P_s = \Psi_{xx} X_s + \Psi_{xz} Z_s \\
   Q_s = \Psi_{zx} X_s + \Psi_{zz} Z_s \\
   P_t = \Psi_{xx} X_t + \Psi_{xz} Z_t \\
   Q_t = \Psi_{zx} X_t + \Psi_{zz} Z_t 
  \end{array}      
  \right .   
\end{equation}
which lead to two independent $2\times 2$ systems needed to solve for $(\Psi_{xx},\Psi_{xz},\Psi_{zx},\Psi_{zz})$:

\begin{equation}
  \left[
  \begin{array}{cc}
   X_t & Z_t \\
   X_s & Z_s \\
  \end{array}      
  \right]   
  \left(
  \begin{array}{c}
  \Psi_{xx} \\
  \Psi_{xz}
  \end{array}
  \right)
  = 
  \left(
  \begin{array}{c}
  P_t \\
  P_s
  \end{array}
  \right)
\end{equation}

\begin{equation}
  \left[
  \begin{array}{cc}
   X_t & Z_t \\
   X_s & Z_s \\
  \end{array}      
  \right]   
  \left(
  \begin{array}{c}
  \Psi_{zx} \\
  \Psi_{zz}
  \end{array}
  \right)
  = 
  \left(
  \begin{array}{c}
  Q_t \\
  Q_s
  \end{array}
  \right)
\end{equation}
These systems have a unique solution under the assumption that the map $(s,t)\to(X,Z)$ is invertible. Using Cramer's rule, we reach an  expression for the Laplacian:
\begin{equation}\label{delta_psi_eq}
  \Delta \Psi = \frac{P_t Z_s - Z_t P_s + X_t Q_s - Q_t X_s}{X_t Z_s - Z_t X_s}
\end{equation}

Denoting the Jacobian of the ray-mapping by
\begin{equation}
  J(s,t) = x_t z_s + x_s z_t
\end{equation}
we note that \eqref{delta_psi_eq} can also be expressed as
\begin{equation}\label{delta_psi_eq_jac}
  \Delta \Psi = \frac{J_t}{J}
\end{equation}
This implies that an expression for the amplitude can be given in terms of the Jacobian:
\begin{equation}
 A(s,t) = \frac{J(s,t)}{J(s,0)} A(s,0)
\end{equation}

can be obtained by adding extra derivatives with respect to $s$ to the ODE system.

\bibliographystyle{plain}
\bibliography{refs}

\end{document}